\documentclass[superscriptaddress,twocolumn,showpacs,prb,amsmath,amssymb]{revtex4}

\usepackage{bm}
\usepackage{bbm}
\usepackage{graphicx}

\begin{document}

\title{Boundary conditions for spin diffusion}

\author{Victor~M.~Galitski} \affiliation{Department of Physics, University of Virginia,
Charlottesville, VA 22904-4714}

\author{Anton~A.~Burkov} \affiliation{Department of Physics, Harvard University, Cambridge, MA 02138}

\author{Sankar~Das~Sarma} \affiliation{Condensed Matter Theory Center, Department of Physics, University of Maryland,
College Park, Maryland 20742-4111}

\begin{abstract}
We develop a general scheme of deriving boundary conditions for
spin-charge coupled transport in disordered systems with spin-orbit
interactions. To illustrate the application of the method, we
explicitly derive boundary conditions for spin diffusion in the
Rashba model. Due to the surface spin precession, the boundary
conditions are non-trivial and contain terms, which couple different
components of the spin density.  We argue that boundary conditions
and the corresponding electric-field-induced spin accumulation
generally depend on the nature of the boundary and therefore the
spin Hall effect in a spin-orbit coupled system can be viewed as a
non-universal edge phenomenon.
\end{abstract}


\maketitle

\section{Introduction}
{There has been a lot of recent interest in a phenomenon dubbed
the spin Hall
effect~\cite{sinova_intrinsic,DP_extrinsic,Zhang_science}.
Experimentally this effect manifests itself as an equilibrium spin
accumulation near the edges, when an electric field is applied
parallel to a Hall bar of a spin-orbit  coupled electron system.
This kind of electric-field-induced spin polarization has been
observed in recent experiments~\cite{exp1,exp2}. This phenomenon
has a potential for being technologically important, as it allows
one to electrically manipulate the spin degree of freedom in
non-magnetic systems. There are still many controversies regarding
the physical origin of the spin accumulation observed in real
experiment (intrinsic~\cite{sinova_intrinsic} {\em vs.}
extrinsic~\cite{DP_extrinsic,Tse_SDS}). But even within a given
theoretical model ({\em e.g.}, a two-dimensional electron gas with
Rashba or Dresselhaus spin-orbit interactions or Luttinger model),
there is still no consensus on the correct description of the spin
Hall effect~\cite{sinova_review}.

A common theoretical approach to the intrinsic spin Hall effect is
as follows: One introduces the notion of a spin current, which is an
operator intuitively related to spin transport.  Using the Kubo
formula, one then calculates the spin-current autocorrelator and the
corresponding linear response quantity, called the ``spin Hall
conductivity''~\cite{Zhang_science,sinova_intrinsic}. In clean
systems, it appears to be related to an elegant topological Berry's
phase structure~\cite{Berry1,Zhang_science}. In a disordered system,
the notion of the Fermi surface Berry's phase is unclear, but the
``spin Hall conductivity'' is well-defined. It is believed that this
 quantity is connected to spin accumulation in the spin-Hall experiment. While this  theoretical approach is mathematically
well-defined and may lead to technically challenging
problems~\cite{SMH}, its relation to experiment remains unclear.
The problem is that in spin-orbit coupled systems, the spin is not
conserved and thus the definition of the spin current is
ambiguous~\cite{Rashba_currents,currents}. A non-equilibrium spin
density in a region relaxes not only due to the flux through the
boundaries of the region, but also due to spin precession.
Therefore, the spin current is not easily measurable.

To describe realistic experiment, it is preferable to use another
route without involving the notion of a spin current. In
disordered systems, this can be done with the help of a kinetic
equation~\cite{MSH} or a spin-charge coupled diffusion
equation~\cite{burkov,Malshukov}. The diffusion equation provides
a complete and physically meaningful description of spin and
charge diffusive transport in terms of position and time dependent
spin and charge densities. However, the explicit solution of the
diffusion equation (which is generally a complicated set of
partial differential equations) strongly depends on boundary
conditions ~\cite{Blei}. While for usual charge transport,
boundary conditions are often obvious and follow from conservation
laws, this is not the case in spin-orbit coupled systems. The
problem is that the spin is not conserved and thus there is spin
precession at the edges. This spin dynamics is very sensitive to
the actual boundary (depending on the physical situation, the
boundary could be modelled as a hard wall, WKB potential, rough
surface, {\em etc.}). Since the bulk diffusion equation describes
spin/charge dynamics only at the length-scales much larger than
the mean-free path $l$, it can not capture the behavior of the
spin in the immediate vicinity of the sample edges. Due to the
aforementioned boundary spin precession, the ``trivial boundary
conditions'' ({\em i.e.}, boundary conditions, which one would
have, if the spin had been conserved) acquire corrections
proportional to the spin-orbit coupling. These corrections are
very important, as they select a unique solution of the diffusion
equation and determine the observable spin accumulation.

The derivation of boundary conditions is technically a difficult
task. Generally, one needs to determine the behavior of the system
in the ballistic region near the boundary and match it with the
diffusive behavior in the bulk. It requires the knowledge of the
exact $S$-matrix or the boundary Green's function. We should point
out that there exists an extensive literature on the related
issues in the context of the diffusion of light in a random
medium~\cite{light_diff} and neutron diffusion \cite{neutrons}. In
these problems, the general form of boundary conditions is obvious
due to conservation laws and only numerical coefficients are
unknown. There exists a general method of deriving these numerical
coefficients. But there are just a very few models in which exact
analytical results are available (a notable example is the Milne
problem~\cite{light_diff}, which describes a boundary separating
diffusive and ballistic regions). We note that in the context of
spin diffusion in spin-orbit coupled systems, even the qualitative
form of boundary conditions is not known as there are no obvious
conservation laws, which would provide guidance. Even though in
the bulk, one has conserved quantities labeled by the band index
({\em e.g.}, chirality in the Rashba model), the boundary
generally does not respect these conservation laws and mixes up
different bands.

In this paper we formulate a general method of deriving boundary
conditions in spin-orbit coupled systems. Using this proposed
method, we derive boundary conditions for the Rashba
model~\cite{Rashba} in the leading order with respect to the
spin-orbit coupling, which is assumed small. We show that there
are corrections to the ``trivial boundary conditions:'' In leading
order, we find that at the boundary, the spin component
perpendicular to the plane of the two-dimensional gas gets coupled
to the  in-plane spin component perpendicular to the boundary. We
argue that  spin accumulation in the spin Hall effect setup is
determined by the combination of the bulk spin-charge coupling and
boundary effects.

\section{General method of deriving boundary conditions}

In this section we formulate a general method of deriving boundary
conditions applicable to any  spin-orbit coupled system with a
boundary. Let us consider a disordered electron system occupying a
region ${\cal A}$.
One can derive the following
integral equation for the diffuson~\cite{Altshuler_Aronov}, which
determines the dynamics of the charge density ($\rho_0$) and the
spin densities ($\rho_\alpha$, $\alpha=x,y,z$):
\begin{eqnarray}
\label{de} \nonumber {\cal D}_{ab} (\omega;{\bf r}_1,{\bf r}_2) =
&& \!\!\!\!\!\! \sum\limits_c \int\limits_{{\bf r}' \in \cal A}
\frac{\Pi_{ac} (\omega;{\bf r}_1,{\bf r}')}{2 \pi \nu \tau} {\cal D}_{cb} (\omega;{\bf r}',{\bf r}_2) d^d{\bf r}'\\
&&  \!\!\!\!\!\!  +\delta_{ab} \delta({\bf r}_1 - {\bf r}_2),
\end{eqnarray}
where $\nu$ is the density of states per spin at the Fermi surface, $\tau$ is the scattering time,
the Latin indices label charge ($0$) and spin ($x$, $y$, and $z$) degrees of freedom,
and ${\cal D}_{ab} (\omega;{\bf r}_1,{\bf r}_2)$ is the diffuson, which is the Green's function of the
spin-charge coupled diffusion equation. The kernel in Eq.~(\ref{de}) reads
\begin{equation}
\label{Pi}
\Pi_{ab} (\omega;{\bf r},{\bf r}') = \frac{1}{2} {\rm Tr}\, \left\{ \hat{\sigma}_a \hat{\tilde{G}}^R (\varepsilon_1;{\bf r},{\bf r}')
\hat{\sigma}_b \hat{\tilde{G}}^A (\varepsilon_2;{\bf r}',{\bf r}) \right\},
\end{equation}
where $\omega = \varepsilon_1 - \varepsilon_2$, $\hat{\sigma}_0$ is the unity matrix, $\hat{\sigma}_\alpha$ are the Pauli matrices,
and $\hat{\tilde{G}}^{R/A} (\varepsilon;{\bf r},{\bf r}')$ are disorder-averaged
retarded/advanced electron Green's functions, which include the effect of the boundary (here and below we denote the boundary Green's functions
as $\tilde{G}$ and the bulk Green's functions as $G$). They depend exponentially on distances via factors of the following types, $\propto e^{-r/(2l)}$, where $l$
is the mean-free path. Deep in the bulk, the edge effects are exponentially small and the bulk Green's functions
can be used, $G^{R/A} \left(\varepsilon, \left| {\bf r} - {\bf r}' \right| \right)$. Performing a gradient expansion
of the bulk polarizability ({\em i.e.}, taking the limit $l \to 0$), one obtains the spin-charge coupled diffusion equation.
Generally, it has the following form:
\begin{equation}
\label{de2}
{\partial_t \rho_a} = \left( D_a \partial^2 - \tau_a^{-1} \right) \rho_a  + \sum\limits_{ b, \gamma}
\Gamma_{ b \gamma}^a \partial_\gamma \rho_b,
\end{equation}
where $\rho_a({\bf r},t)$ are the charge/spin densities, $D_a$ are the diffusion coefficients, $\tau_a$ are the relaxation times,
and $\Gamma_{b \gamma}^a$ are the spin-spin and spin-charge couplings.

Note that the integral equation (\ref{de}) does not require
boundary conditions (it already contains this information through
the boundary Green's function), but the differential diffusion
equation (\ref{de2}) does. To derive these boundary conditions one
needs to perform the gradient expansion near the boundary
$\partial {\cal A}$. Let us consider the problem in half-space
${\cal A} = \left\{x \left| \right. x > 0 \right\}$. To find the
diffusion equation near the boundary, we take the limit $x = {\bf
r} \cdot {\bf n} \to 0$. This limit should be understood in the
following sense: $ p_{\rm F}^{-1} \ll x \ll l$, {\em i.e.}, we
consider a point, which is near the boundary as compared to the
diffusion length-scale $l$, but still far from it as compared to
the ballistic length scale described by the Fermi wave-length. Let
us first consider the case of zero spin-orbit interactions. If the
boundary is a hard wall potential, then the boundary Green's
function can be obtained using the method of mirror images
\begin{equation}
\label{mirror}
\hat{\tilde{G}}_0^{R/A} (\varepsilon;{\bf r},{\bf r}') =
\left[ G_0^{R/A} \left(\varepsilon;\Delta r \right) - G_0^{R/A} \left(\varepsilon;\Delta r_*\right)
\right] \hat{\sigma}_0,
\end{equation}
where  $\Delta{\bf r} = {\bf r} - {\bf r}'$, $\Delta {\bm r}_* =
{\bf r}_* - {\bf r}'$, and ${\bf r}_*$ is the mirror image with
respect to the sample boundary. 
For the sake of concreteness,
let us assume that the dimensionality $d=2$ (all qualitative
arguments are independent of dimensionality). In two dimensions, the
Green's function has the following form
\begin{equation}
\label{2DGF0} G_0^{R/A}(\varepsilon; r) = - \sqrt{ \frac{ \pm i
m^2}{2 \pi p_{\rm F} r}} \exp{\left[ \pm i \left(p_{\rm F} +
\frac{\varepsilon}{v_{\rm F}} \right) r - \frac{r}{2l} \right]}.
\end{equation}
We note here that the frequency dependence of the diffusion kernel is irrelevant for the boundary condition problem and will be omitted from
now on. When the method of mirror images applies, the polarizability kernel (\ref{Pi}) has the following general structure
$\Pi^{(0)} = G^R G^A + G_*^R G_*^A - G^R G_*^A - G_*^R G^A  $ [where $G_*$ implies the ``reflected'' Green's function, {\em i.e}, the second term in Eq.~(\ref{mirror})].
The key observation is that due to the fast-oscillating factor
$GG_* \propto\exp\left[\pm i p_F \left( \Delta r - \Delta r_* \right)\right]$ and our choice of the point ${\bf r}$ ($x \gg p_{\rm F}^{-1}$),
 the cross terms $GG_*$ in the expression for $\Pi^{(0)}$ become negligible in the integral over ${\bf r}'$  and can be omitted.
The remaining integral can be easily calculated and we find (in
the ``trivial'' case of no spin-orbit coupling):
\begin{equation}
\label{Pi0}
\lim\limits_{{\bf r} \to {\bf 0}} \sum\limits_b \int\limits_{x' > 0}
\frac{\Pi_{ab}^{(0)} ({\bf r},{\bf r}')}{2 \pi \nu \tau} \rho_b({\bf r}') d^d{\bf r}' = \rho_a({\bf 0}) + c_d l\,
{\partial_x \rho_a({\bf 0})} ,
\end{equation}
where $c_d$ is a dimensionless coefficient, which in two dimensions is equal to $c_2 = 2/\pi$.
From Eqs.~(\ref{de}) and (\ref{Pi0}), we recover the familiar boundary condition
${\bf n} \cdot {\bm \partial} \rho({\bf 0}) = 0$, which is simply the consequence
of the charge and spin conservation.

In a system with spin-orbit interaction, the method of images does
not apply because the ``reflected'' Green's function does not
satisfy the corresponding Schr{\"o}dinger equation. This  is a
major complication, which never occurs in the problems usually
studied in the literature in this
context~\cite{light_diff,neutrons}. In addition, the matrix
structure of the boundary diffusion kernel becomes essential.
There are generally two ways to obtain the boundary Green's
function (we have explicitly verified that both approaches lead to
identical perturbative results in the leading order~\cite{myown}):
(i)~The first is to solve the Schr{\"o}dinger equation near the
boundary and express the Green's function using the corresponding
wave-functions as a basis; (ii)~The second is to find the Green's
functions perturbatively, using the mirror-image result
(\ref{mirror}) as the leading approximation:
\begin{eqnarray}
\label{expansion}
\nonumber
\delta\hat{\tilde{G}} ({\bf r}, {\bf r}') = \sum\limits_{k=1}^{\infty}\!
\int\limits_{x_k > 0}\!\! \ldots \!\!\int\limits_{x_1 > 0} && \!\!\!\!\!\! d^d{\bf r}_1 \! \ldots \! d^d{\bf r}_k
\tilde{G}_0 ({\bf r},{\bf r}_1) \hat{V}_{\rm SO}({\bf r}_1)  \\
&& \!\!\!\!\!\! \!\!\!\!\!\! \!\!\!\!\!\!  \!\!\!\!\!\!
\!\!\!\!\!\!  \!\!\!\!\!\!\!\!
 \times \tilde{G}_0 ({\bf r}_1,{\bf r}_2) \times  \ldots \times  \hat{V}_{\rm SO}({\bf r}_k) \tilde{G}_0 ({\bf r}_k,{\bf r}'),
\end{eqnarray}
where $\hat{V}_{\rm SO}$ is the spin-orbit interaction operator as
it appears in the Hamiltonian. We note that disorder corrections
to the spin-orbit vertex can be shown to be small as long as
$E_{\rm F} \tau \gg 1$. The resulting Green's function
(\ref{expansion}) may be quite complicated even in the simplest
cases. However, the diffusion kernel always contains products of
the following types:
\begin{eqnarray}
\nonumber
 \Pi_{ab} ({\bf r},{\bf r}') \propto && \left[ T G^R_0
(\Delta {\bf r}) - R G^R_0 (\Delta {\bf r}_*) \right]\\
&\times& \left[ T' G^A_0 (\Delta {\bf r}) - R' G^A_0 (\Delta {\bf
r}_*) \right],
\end{eqnarray}
where $T$'s and $R$'s are some functions, which do not oscillate
on a Fermi wavelength scale. Again, in the integral of
Eq.~(\ref{de}), the cross-terms $G_0^{R/A} (\Delta {\bf r})
G^{A/R}_0 (\Delta {\bf r}_*)$ become negligible due to the
fast-oscillations and can be omitted, but the other terms survive.
One can define the following matrix, which determines  boundary
conditions
\begin{equation}
\label{B_matrix}
B_{ab} = \lim\limits_{{\bf r} \to {\bf 0}} \int\limits_{x' > 0}
\frac{\delta\Pi_{ab} ({\bf r},{\bf r}')}{2 \pi \nu \tau}  d^d{\bf r}',
\end{equation}
where the correction to polarizability is determined by the
Green's function (\ref{expansion}) and Eq.~(\ref{Pi}) and the
limit is understood in the sense $p_{\rm F}^{-1} \ll | {\bf r}
\cdot {\bf n} | \ll l$. If the boundary is an impenetrable wall,
general boundary conditions take the form
\begin{equation}
\label{gen_bc}
{\bf n} \cdot {\bm \partial} \rho_a({\bf 0}) = -\frac{1}{c_d l} \sum \limits_b B_{ab} \rho_b ({\bf 0}),
\end{equation}
where $a,b = 0,x,y, \mbox{ and } z$, ${\bf n}$ is a unit vector
normal to the boundary, $c_d$ is a number ($c_2 = 2/\pi$), and
$B_{ab}$ is a dimensionless matrix defined above. If an external
field is present one may need to consider another term in the
boundary conditions, which has the meaning of the boundary
spin-charge coupling~\cite{Bleibaum}
\begin{equation}
\label{bsc}
 {\bf C}_a =  \lim\limits_{{\bf r} \to {\bf 0}}
\int\limits_{x'
> 0} {\delta\Pi_{ab} ({\bf r},{\bf r}')}{\bf r}' d^d{\bf r}'/({2
\pi \nu \tau}). \end{equation} In this case the boundary condition
becomes:
\begin{equation}
\label{bc2}
 {c_d l} {\bf n} \cdot {\bm
\partial} \rho_a({\bf 0}) = - \sum \limits_b B_{ab} \rho_b ({\bf 0})
- {\bf C}_a {\bm\nabla}\rho_0({\bf 0})
\end{equation}

We note that if the boundary is not impenetrable, the general
method does not change but the results do even in the lowest
order. For instance, if the boundary is characterized by an
isotropic reflection coefficient $R \ne 1$, the zeroth order
boundary condition becomes $(1-R^2) \rho_a({\bf 0}) = c_d l
(1+R^2) {\bf n} \cdot {\bm \partial} \rho_a({\bf 0})$. If the
reflection coefficient is not too close to unity, the right-hand
side of the latter equation can be considered small and one can
use the following leading order boundary conditions
\begin{equation}
\label{S=0}
 \rho_a ({\bf 0}) = 0.
\end{equation}
This boundary condition implies that an excess charge or spin
density can not exist at the boundary if the boundary is
penetrable. The excess density leaks through the edge (see also
Ref.~[\onlinecite{bauer}]).

We note that strictly speaking the matching point at the boundary
[which we set to zero in Eqs.~(\ref{bc2}) and (\ref{S=0})]  can
not be obtained exactly within the method described here, but all
dimensional parameters and numerical coefficients are robust. To
determine the correct matching point on a lengthscale of order
$l$, one actually needs to solve exactly the integral equation,
which is possible only in very special cases [{\em e.g.}, the
Milne problem ({\em i.e.}, R=0 and no spin-orbit coupling)].

\section{Application to the Rashba model}

To illustrate the application of our method, let us consider the
Rashba model~\cite{Rashba} in the leading order in spin-orbit
coupling. In this case, the spin-orbit interaction operator has
the form $\hat{V}_{\rm SO} = - i \alpha_{\rm R} \epsilon_{\alpha
\beta z} \hat{\sigma}_\alpha
\partial_\beta$, where $\alpha_{\rm R}$ is the Rashba spin-orbit
coupling and $\alpha,\beta = x,y,z$. The differential diffusion
equation was derived in Ref.~[\onlinecite{burkov}] and reads
\begin{equation}
\label{deR0}
\partial_t \rho_0 = D \partial^2 \rho_0 - \Gamma_{\rm cs} \epsilon_{z\alpha\beta} \partial_\alpha \rho_\beta;
\end{equation}
\begin{equation}
\label{deRxyz}
\partial_t \rho_\alpha = \left(D \partial^2 - \tau_\alpha^{-1} \right) \rho_\alpha +
\Gamma_{\rm ss} \epsilon_{\alpha \beta \gamma} \epsilon_{z \lambda \beta} \partial_\lambda \rho_\gamma -
\Gamma_{\rm sc} \epsilon_{z\beta\alpha} \partial_\beta \rho_0.
\end{equation}
The spin-spin coupling in the Rashba model is $\Gamma_{\rm ss} = 4
\alpha_{\rm R} E_{\rm F} \tau$ and the spin relaxation times are
$2 \tau_z = \tau_x = \tau_y = \tau_{\rm s}= 2\tau/(2 \alpha_{\rm
R} p_{\rm F} \tau)^2$. These parameters are universal in the sense
that they do not depend on the  details of the electronic spectrum
and can be calculated in the $\xi$-approximation~\cite{AGD}.
However, the spin-charge coupling is not universal, since it
strongly depends on the electronic spectrum and the high-energy
cut-off (if the spectrum is quadratic, $\Gamma_{\rm sc} =2
\alpha^3_{\rm R} p_{\rm F}^2 \tau^2$). We should point out however
that this result holds only in the semiclassical diffusion
approximation, {\em i.e.}, only in the leading order with respect
to the inverse conductance $1/ (E_{\rm F} \tau)$, which is an
independent parameter of the model. There may exist contributions
to the spin-charge coupling  of order $\Gamma_{\rm sc} \sim
\Gamma_{\rm ss}/ (E_{\rm F} \tau)$. They originate from diagrams
with crossed impurity lines, which are omitted in the diffusion
approximation (we note that there is no direct relation between
$\Gamma_{\rm sc}$ and the spin Hall conductivity, which vanishes
to all orders~\cite{Dimitrova}). These quantum corrections can be
neglected only if $\left(E_{\rm F} \tau\right)^{-1/2} \ll \alpha
p_{\rm F} \tau \ll 1$. This is a very strong constraint, which is
not always satisfied in experimentally studied systems. In what
follows, we will treat the spin-charge coupling terms in the
diffusion equation (\ref{deR0},~\ref{deRxyz}) as non-universal
phenomenological parameters and assume that the above-mentioned
constraint is satisfied.

The diffusion equation (\ref{deR0},~\ref{deRxyz}) must be
supplemented by boundary conditions. We consider the problem in
leading approximation with respect to the spin-orbit coupling
parameter $\alpha_{\rm R}$ and the inverse conductance. First, we
use Eqs.~(\ref{Pi}) and (\ref{expansion}) to derive a correction
to the polarizability kernel
\begin{equation}
\label{Pab}
\delta \Pi_{\alpha\beta}^{(1)} ({\bf r},{\bf r}') = - 2 \alpha_{\rm R} \epsilon_{\alpha\beta\gamma}\epsilon_{\gamma\delta}
{\rm Re\,} \left[ \tilde{G}_0^A ({\bf r},{\bf r}') g_\delta ({\bf r},{\bf r}') \right]
\end{equation}
and
\begin{equation}
\label{P0b}
\delta \Pi_{0 \beta}^{(1)} ({\bf r},{\bf r}') = \delta \Pi_{ \beta 0}^{(1)} ({\bf r},{\bf r}')= 2 \alpha_{\rm R} \epsilon_{\beta\delta}
{\rm Im\,} \left[ \tilde{G}_0^A ({\bf r},{\bf r}') g_\delta ({\bf r},{\bf r}') \right],
\end{equation}
where the boundary Green's functions are defined by
Eqs.~(\ref{mirror}) and (\ref{2DGF0}) and we introduced the
following function
$$
g_\alpha ({\bf r},{\bf r}') =
\int\limits_{x_1>0} d^2{\bf r}_1 \tilde{G}_0^R ({\bf r},{\bf r}_1)
\partial_{1\alpha} \tilde{G}_0^R ({\bf r}_1,{\bf r}').
$$
This function contains four terms (we can symbolically denote them
as $g_{1\alpha}= \int G \partial_\alpha G$, $g_{2\alpha}=-\int G_*
\partial_\alpha G$, $g_{3\alpha}=-\int G \partial_\alpha G_*$, and
$g_{4\alpha}=\int G_* \partial_\alpha  G_*$). Let us analyze the
first one, the other terms can be calculated in a similar manner:
To calculate the corresponding integral, we use a new coordinate
system $(\tilde{x},\, \tilde{y})$ in which the points ${\bf r}$ and
${\bf r}'$ have the coordinates $(0,-\Delta r/2)$ and $(0,\Delta
r/2)$ respectively ($\Delta {\bf r} = {\bf r - r}'$). Next, we
introduce the elliptic coordinates $(\eta,\tilde{\phi})$, with the
points ${\bf r}$ and ${\bf r}'$ in the foci of the ellipses: $|{\bf
r}_1 - {\bf r}| + |{\bf r}' - {\bf r}_1| = \eta \Delta r$ and
$\tilde{\phi}$ being the angle between the vector ${\bf r}_1$ and
the $\tilde{x}$ axis. Using these elliptic coordinates and
Eq.~(\ref{2DGF0}), we write the integral  $g_{1\alpha}$ in the
following form
\begin{eqnarray}
\nonumber
 g_{1 \alpha} ({\bf r},{\bf r}') =  - \frac{i
m^2}{4 \pi p_{\rm F}} \partial'_{\alpha} \Biggl[ \Delta r \!\!
\int\limits_1^\infty d\eta  \!\! \int\limits_{x_1>0}
\!\!\!\!\!\!\!\!\!\!\!\! && d\tilde{\phi}\,\sqrt{ \frac{ \eta^2 -
\sin^2{\tilde{\phi}}}{\eta^2 -1}}  \\ &\times& e^{i \left( p_{\rm
F} + \frac{i}{2l} \right) \eta \Delta r } \Biggr].
\label{el_coord}
\end{eqnarray}
We note that the limits of integration over $\tilde{\phi}$ are
generally non-trivial due to the constraint $x_1 > 0$. It is
possible to evaluate analytically the correction to the diffusion
kernel [see Eqs.~(\ref{Pi}), (\ref{Pab}), and (\ref{P0b})] and the
boundary matrix (\ref{B_matrix}). Indeed, from
Eqs.~(\ref{B_matrix}), (\ref{Pab}), and (\ref{P0b}), we find that
due to the fast-oscillating exponents, only a few terms survive the
integration over ${\bf r}'$. These remaining terms contain
combinations of the following types $\exp\left[ i p_{\rm F}  \Delta
r (\eta -1 ) \right]$, which constrain the parameter $\eta$ in
(\ref{el_coord}) to be close to unity. We find in the leading order
the following results for the boundary matrix~(\ref{B_matrix})
\begin{equation}
\label{Bob}
B_{\alpha \beta} = \frac{4 \alpha_{\rm R} m l}{\pi} \epsilon_{\alpha \beta y}\,\, \mbox{ and } \,\,
B_{0 b} = B_{b 0} = \delta_{by} {\cal O}\left( \frac{\alpha_{\rm R} m l} {E_{\rm F} \tau} \right),
\end{equation}
where the last equation implies that the coupling between the spin
and charge densities is small with respect to the inverse
conductance and vanishes in the framework of the diffusion
approximation and the accuracy of the method.

From  Eqs.~(\ref{gen_bc}) and (\ref{Bob}), we find the following
boundary conditions in the leading approximation with respect to
the spin-orbit coupling and the inverse conductance:
\begin{equation}
\label{c}
 \partial_x \rho_0 ({\bf 0}) = \partial_x \rho_y ({\bf 0}) = 0,
\end{equation}
\begin{equation}
\label{x}
\partial_x \rho_x ({\bf 0}) = - \frac{\Gamma_{\rm ss}}{2 D} \rho_z({\bf 0}),\,\,\,\mbox{and}\,\,\,
\partial_x \rho_z ({\bf 0}) =  \frac{\Gamma_{\rm ss}}{2 D} \rho_x({\bf 0}),
\end{equation}
where $\Gamma_{\rm ss}$ is defined after Eq.~(\ref{deRxyz}). These
boundary conditions can be generalized to describe any form of a
hard-wall boundary.
To find such a general form of boundary conditions, one should do
the following replacements: $\partial_x \to {\bf n} \cdot {\bm
\partial}$ and  $\rho_x \to \rho_{\bf n}$, where ${\bf n}$ is the
unit vector normal to the segment of the boundary and $\rho_{\bf
n}$ is the corresponding component of the spin density. However,
if the boundary is a smooth potential with the length-scale in the
${\bf n}$-direction being $a$, then boundary conditions would be
different and strongly dependent on the ratios of $a/l$ and
$ap_{\rm F}$~\cite{myown}. Also, the boundary potential would lead
to an additional boundary spin-orbit coupling, which would
strongly affect the boundary conditions.

We note here that in the next-to-leading order with respect to the
spin-orbit interaction, two new effects appear: Boundary
spin-charge coupling and boundary spin relaxation (the former is
parametrically larger than the latter due to an additional
smallness of the spin density). To this order boundary conditions
take the form:
$$
{\bf n} \cdot {\bf
\partial} \rho_z ({\bf 0}) = \frac{\Gamma_{\rm ss}}{2 D} \rho_n
({\bf 0}) + \gamma \rho_z({\bf 0}) - {\bf C}_z {\bf
\nabla}\rho_0({\bf 0}),
$$
where $\gamma$ and ${\bf C}_z$ are constants $\propto \alpha_{\rm
R}^2$. We should note that the boundary spin-charge coupling term
can be shown to be strongly dependent on the electronic spectrum
and the physics far from the Fermi line and as such is
non-universal and can not be calculated in the $\xi$-approximation
used in this paper.


\section{Solution of the diffusion equation}

We now solve the diffusion equation for a two-dimensional gas of
Rashba electrons occupying a half-space $x>0$ and in the presence
of an electric field parallel to the edge. We consider two types
of edges:  a partially or totally transparent interface and an
impenetrable boundary. It is convenient to rewrite the diffusion
equation (\ref{deR0},~\ref{deRxyz}) in terms of the rescaled
variables: $\bar{\rho}_x(\bar{x}) = [\rho_x(\bar{x}) -
\rho_\infty]/\rho_\infty$ and $\bar{\rho}_z(\bar{x}) =
\rho_z(\bar{x})/\rho_\infty$, where $\bar{x} = x/\sqrt{D \tau_{\rm
s}}$ is the dimensionless distance and $\rho_\infty =  2
\Gamma_{\rm sc} \nu \tau_{\rm s} e E$ is the uniform spin density
due to the spin-charge coupling. In these variables, the diffusion
equation takes the following form
\begin{equation}
\label{d1}
\bar{\rho}_x'' - \bar{\rho}_x + 2\bar{\rho}_z' = 0;
\end{equation}
\begin{equation}
\label{d2} \bar{\rho}_z'' - 2\bar{\rho}_z - 2\bar{\rho}_x' = 0.
\end{equation}
Boundary conditions for a transparent boundary are
$\bar{\rho}_z(0)=0$ and $\bar{\rho}_x(0) =-1$ and for a hard-wall
boundary $\bar{\rho}_x'(0) = - \bar{\rho}_z(0)$ and
$\bar{\rho}_z'(0) = \bar{\rho}_x(0) + \zeta$, where $\zeta = 1 -
C_{zy} \partial_y \rho_0/\rho_\infty$ is a non-universal
spectrum-dependent parameter. We note that if the latter is zero,
one has to take into account the next-to-leading order boundary
spin relaxation terms, which should determine the observable spin
density. The solution of the differential
equation~(\ref{d1},~\ref{d2}) is straightforward and we find for
the transparent boundary (e.g., ballistic contacts)
\begin{equation}
\label{s1} \bar{\rho}_x(\bar{x}) = \frac{-1}{{\rm Re}\,
\left(\frac{k}{k^2+1}\right)} {\rm Re}\, \left( \frac{k
e^{ik\bar{x}}}{k^2+1} \right)
\end{equation}
and
\begin{equation}
\label{s12} \bar{\rho}_z(\bar{x}) =- \frac{{\rm Im}\, \left(
e^{ik\bar{x}} \right)}{2{\rm Re}\, \left(\frac{k}{k^2+1}\right)}
\end{equation}
and for the hard-wall boundary
\begin{equation}
\label{s2} \bar{\rho}_x(\bar{x}) = \frac{2 \zeta}{{\rm Re\,} k}
{\rm Re\,} \left[ \frac{k e^{i k \bar{x}} }{k^2-1} \right]
\end{equation}
 and
\begin{equation}
\label{s22}
 \bar{\rho}_z(\bar{x}) = \frac{\zeta}{{\rm
Re\,} k} {\rm Im\,} \left[ \frac{k^2+1}{k^2-1} e^{i k \bar{x}}
\right],
\end{equation}
where $k = \sqrt{(1+i\sqrt{7})/2} \approx 0.978 + 0.676 i$ is an
eigenvalue of the diffusion equation~(\ref{d1},~\ref{d2}). Since
the corresponding differential operator is non-Hermitian, the
eigenvalues are complex. This leads to oscillations of the spin
density near the boundary with the period $l_{\rm osc} \approx
6.422 \sqrt{D \tau_{\rm s}}$. We should point out that there
definitely exists other types of small-lengthscale oscillations
(with the period $\pi/p_{\rm F}$), which are superimposed onto the
regular behavior (\ref{s1} -- \ref{s22}). However, these Friedel
oscillations are beyond the reach of the diffusion approximation.
From Eqs.~(\ref{s1} -- \ref{s22}), we see that the spin density
rapidly decays as we move away from the boundary. The
corresponding lengthscale is $l_{\rm s} \approx 1.479 \sqrt{D
\tau_{\rm s}}$.

\section{Summary}

In this paper, we have developed a general method of deriving
boundary conditions for spin diffusion equation. The method
involves a gradient expansion of the boundary Green's function,
which takes into account the behavior of the electron
wave-functions on ballistic lengthscales near the edges. Using the
proposed method, we have found a general form of boundary
conditions in the Rashba model for two types of edges: a
transparent interface and a hard wall.

We have found an exact solution of the diffusion equation
satisfying these two types of boundary conditions. The spin
accumulation (\ref{s1} -- \ref{s22}) is shown to oscillate and
decay away from the boundary with the corresponding lengthscales
being determined by the eigenvalues of the bulk diffusion
equation. However, we argue that even small changes in the
boundary potential would lead to a different solution of the
diffusion equation: The eigenvalues, which determine the density
profile, are robust and universal, but the overall amplitude and
phase shifts are boundary-specific. Therefore, the spin Hall
effect is generally a non-universal phenomenon, which depends on
the structure of the sample edges.

The authors are grateful to L. Balents, H.-A.Engel,  A.G. Malshukov,
E.I. Rashba, and Ya. Tserkovnyak for helpful discussions. A.A.B. is
supported by the NSF under grant DMR02-33773. S.D.S. is supported by
US-ONR, NSF, and LPS.

\vspace*{-0.2in}
}
\bibliography{bc}

\begin{thebibliography}{24}
\bibitem[{Ble()}]{Bleibaum}
\bibinfo{howpublished}{The boundary spin-charge coupling in the Rashba model
  was discussed separately by O.~Bleibaum in cond-may/0606351.}
\end{thebibliography}

\end{document}